\documentclass[12pt,preprint]{aastex}

\newcommand{\apg}{\gtrsim}
\newcommand{\apl}{\lesssim}
\newcommand{\cmjj}{\mbox{${\rm cm^{-2}}$}}

\newcommand{\kms}{\mbox{km\ s${^{-1}}$}}
\newcommand{\lya}{\mbox{${\rm Ly}\alpha$}}

\begin{document}

\title{THE STAR FORMATION RATE INTENSITY DISTRIBUTION FUNCTION---IMPLICATIONS
FOR THE COSMIC STAR FORMATION RATE HISTORY OF THE UNIVERSE$^{1,2,3,4}$}

\author{KENNETH M. LANZETTA, NORIAKI YAHATA, SEBASTIAN PASCARELLE \\
Department of Physics and Astronomy, State University of New York at Stony
Brook \\
Stony Brook, NY 11794--3800, U.S.A.}

\author{HSIAO-WEN CHEN \\
Observatories of the Carnegie Institute of Washington \\
813 Santa Barbara St., Pasadena, CA, 91101, U.S.A.}

\and

\author{ALBERTO FERN\'{A}NDEZ-SOTO$^5$ \\
Osservatorio Astronomico di Brera \\
Via Bianchi 46, Merate (LC). I--23807, ITALY}

\altaffiltext{1}{Based on observations with the NASA/ESA Hubble Space
Telescope, obtained at the Space Telescope Science Institute, which is operated
by the Association of Universities for Research in Astronomy, Inc., under NASA
contract NAS5--26555.}

\altaffiltext{2}{Based on observations made at the Kitt Peak National
Observatory, National Optical Astronomy Observatories, which is operated by the
Association of Universities for Research in Astronomy, Inc.\ (AURA) under
cooperative agreement with the National Science Foundation.}

\altaffiltext{3}{Based on observations collected at the European Southern
Observatory, Paranal, Chile (VLT-UT1 Science Verification Program).}

\altaffiltext{4}{Observations have been carried out using the ESO New
Technology Telescope (NTT) at the La Silla observatory under Program-ID Nos.
61.A-9005(A), 162.O-0917, 163.O-0740, 164.O-0561.}

\altaffiltext{5}{Marie Curie Fellow.}

\newpage

\begin{abstract}

  We address the effects of cosmological surface brightness dimming on
observations of faint galaxies by examining the distribution of ``unobscured''
star formation rate intensities versus redshift.  We use the star formation
rate intensity distribution function to assess the ultraviolet luminosity
density versus redshift, based on our photometry and photometric redshift
measurements of faint galaxies in the HDF and the HDF--S WFPC2 and NICMOS
fields.  We find that (1) previous measurements have missed a dominant fraction
of the ultraviolet luminosity density of the universe at high redshifts by
neglecting cosmological surface brightness dimming effects, which are important
at redshifts larger than $z \approx 2$, (2) the incidence of the highest
intensity star forming regions increases monotonically with redshift, and (3)
the ultraviolet luminosity density plausibly increases monotonically with
redshift through the highest redshifts observed.  By measuring the spectrum of
the luminosity density versus redshift, we also find that (4) previous
measurements of the ultraviolet luminosity density at redshifts $z < 2$ must be
reduced by a factor $\approx 2$ to allow for the spectrum of the luminosity
density between rest-frame wavelengths 1500 and 2800 \AA.  And by comparing
with observations of high-redshift damped \lya\ absorption systems detected
toward background QSOs, we further find that (5) the distribution of star
formation rate intensities matches the distribution of neutral hydrogen column
densities at redshifts $z \approx 2$ through 5, which establishes a
quantitative connection between high-redshift galaxies and high column density
gas and suggests that high-redshift damped \lya\ absorption systems trace lower
star formation rate intensity regions of the same galaxies detected in star
light in the HDF and HDF--S.  Because our measurements neglect the effects of
obscuration by dust, they represent {\em lower limits} to the total star
formation rate density.

\end{abstract}
 
\keywords{cosmology:  observations; galaxies:  evolution}

\newpage

\section{INTRODUCTION}

  The rest-frame ultraviolet luminosity density of the universe traces the
history of cosmic star formation because it is produced by hot, massive, young
stars.  Previous measurements have found that the ultraviolet luminosity
density increases with redshift to a maximum at redshift $z \approx 1$ to 2 and
then decreases (Madau et al.\ 1996) or remains constant (Madau, Pozzetti, \&
Dickinson 1998; Steidel et al.\ 1999; Hopkins, Connolly, \& Szalay 2000) with
redshift to higher redshifts.  These previous measurements have been
interpreted in the context of galaxy formation and evolution scenarios as
indicating that the stellar content of galaxies was formed gradually, over most
of cosmic time (e.g.\ Somerville, Primack, \& Faber 2001).

  But previous measurements have neglected cosmological surface brightness
dimming effects, which can play a dominant role in setting what is observed at
high redshifts (Pascarelle, Lanzetta, \& Fern\'andez-Soto 1998).  Specifically,
surface brightness (per unit frequency interval) decreases with redshift as $(1
+ z)^{-3}$ due to the expansion of the universe.  This has an important
consequence for observations of distant galaxies:  while both intrinsically
bright and intrinsically faint regions of low-redshift galaxies are accessible
to observation, only intrinsically bright regions of high-redshift galaxies are
accessible to observation---intrinsically faint regions of high-redshift
galaxies are of low surface brightness and are simply not detected against the
background noise.  Thus all measurements {\em miss} some fraction of the light
of distant galaxies; this fraction is small for low-redshift galaxies but can
be dominant for high-redshift galaxies.

  Here we address this issue by examining the distribution of ``unobscured''
star formation rate intensities versus redshift---i.e.\ the star formation rate
intensities that are directly inferred from the observed rest-frame ultraviolet
light that is not obscured by intervening dust.  We use the star formation rate
intensity distribution function to assess the ultraviolet luminosity density
versus redshift, based on our photometry and photometric redshift measurements
of faint galaxies in the Hubble Deep Field (HDF) and Hubble Deep Field South
(HDF--S) WFPC2 and NICMOS fields (Lanzetta, Yahil, \& Fern\'andez-Soto 1996;
Fern\'andez-Soto, Lanzetta, \& Yahil 1999; Yahata et al.\ 2000; Yahata et al.\
2001 in preparation).  The star formation rate intensity distribution function
bears on the issue because it makes explicit the effects of cosmological
surface brightness dimming on observations of high-redshift galaxies, thereby
allowing the amount by which previous measurements have underestimated the
ultraviolet luminosity density at high redshifts to be directly estimated.
Because our measurements neglect the effects of obscuration by dust, they
represent {\em lower limits} to the total star formation rate density.

  The framework of the unobscured star formation rate intensity and the star
formation rate intensity distribution function is presented in \S\ 2.  The
photometry and photometric redshift measurements are summarized in \S\ 3, the
angular area versus depth relations are summarized in \S\ 4, and the Stony
Brook faint galaxy redshift survey is summarized in \S\ 5.  The star formation
rate intensity distribution function measurements are described in \S\ 6,
results are described in \S\ 7, and the connection with the neutral hydrogen
column density distribution function is described in \S\ 8.  The summary and
conclusions are presented in \S\ 9.  Unless otherwise stated, we adopt a
standard Friedmann cosmological model of dimensionless Hubble constant $h_{100}
= H_0 / (100 \ \kms \ {\rm Mpc}^{-3}$) and deceleration parameter $q_0 = 0.5$;
this is the same cosmological model adopted for previous measurements of the
rest-frame ultraviolet luminosity density.

\section{FRAMEWORK}

  To address the issue of the effects of cosmological surface brightness
dimming, we examine the distribution of ``unobscured'' star formation rate
intensities versus redshift.  We designate the unobscured star formation rate
intensity as $x$ and the star formation rate intensity distribution function
as $h(x)$.  The star formation rate intensity distribution function $h(x)$
bears on the issue because it makes explicit the effects of cosmological
surface brightness dimming on observations of high-redshift galaxies.

  By ``unobscured star formation rate intensity,'' we mean the star formation
rate intensity that is {\em directly inferred} from the rest-frame ultraviolet
light that is not obscured by intervening dust.  Hence unobscured star
formation rate intensity (measured, say, in units of $M_\odot$ yr$^{-1}$
kpc$^{-2}$) is {\em equivalent to} rest-frame ultraviolet luminosity per unit
area (measured, say, in units of erg s$^{-1}$ Hz$^{-1}$ kpc$^{-2}$), once some
fixed scaling between ultraviolet luminosity and star formation rate (which is
set by the stellar initial mass function) is specified.  To determine the {\em
total} star formation rate intensity from the unobscured star formation rate
intensity would require knowledge of the effects of obscuration by intervening
dust, which is beyond the scope of the present analysis.  Rather, here we
address {\em only} the effects of cosmological surface brightness dimming.  We
emphasize that because our measurements neglect the effects of obscuration by
dust, they represent {\em lower limits} to the total star formation rate
density.

  We define the star formation rate intensity distribution function $h(x)$ in
such a way that $h(x) dx$ is the projected proper area per comoving volume of
unobscured star formation rate intensity in the interval $x$ to $x + dx$.  The
star formation rate intensity distribution function $h(x)$ is a fundamental
statistical description of the evolving galaxy population, similar in spirit to
the galaxy luminosity function but cast in terms of star formation rate
intensity (derived from measurement of surface brightness) rather than
luminosity (derived from measurement of energy flux).  For our purposes, the
first moment of $h(x)$
\begin{equation}
\int_0^\infty x \ h(x) \ dx
\end{equation}
is of particular interest.  The first moment of $h(x)$ is the unobscured star
formation rate density $\dot{\rho}_S$, which is equivalent to within a factor
of scale to the ultraviolet luminosity density $\rho_\nu$.  Equation (1) makes
explicit the effects of cosmological surface brightness dimming on measurement
of the ultraviolet luminosity density:  In equation (1), the limits of
integration extend from $x = 0$ through $\infty$.  But in practice, a given
observation (at a given redshift) is sensitive only to some star formation rate
intensity threshold $x_{\rm min}$, which is set by the surface brightness
threshold of the observation (and the redshift).  Whether or not a given
observation (at a given redshift) is suitable for measuring the ultraviolet
luminosity density depends on whether or not $x_{\rm min}$ is less than the
value of $x$ above which the bulk of the ultraviolet luminosity density is
emitted.  Here we assess the ultraviolet luminosity density versus redshift and
determine the amount by which previous measurements have underestimated the
ultraviolet luminosity density at high redshifts, based on our photometry and
photometric redshift measurements of faint galaxies in the HDF and HDF--S WFPC2
and NICMOS fields.

\section{PHOTOMETRY AND PHOTOMETRIC REDSHIFT MEASUREMENTS}

\subsection{Photometry}

  Details of our photometry of faint galaxies in the HDF and HDF--S WFPC2 and
NICMOS fields have been described previously by Fern\'andez-Soto et al.\ (1999)
and Yahata et al.\ (2000).  In this section, we summarize important aspects of
our photometry techniques.

  We processed all available space- and ground-based optical- and
infrared-wavelength images of the HDF and HDF--S WFPC2 and NICMOS fields.
These images include discretionary, GO, and GTO observations obtained with the
Hubble Space Telescope (HST) and images obtained with the Kitt Peak National
Observatory (KPNO) 4 m telescope and the European Southern Observatory (ESO)
New Technology Telescope (NTT) and Very Large Telescope UT--1 (VLT UT--1).  We
registered the images (to an accuracy of typically 0.25 pixel for space-based
images and 0.5 pixel for ground-based images) by appropriate translation,
rotation, and scaling using our own algorithms and software, and we identified
and masked cosmic ray events using our own algorithms and software.  Details of
the observations are summarized in Table 1.

  We detected objects in the images at multiple bandpasses using the SExtractor
program (Bertin \& Arnouts 1996), starting at shorter wavelengths and working
toward longer wavelengths.  First, we detected objects in the images of a
fiducial bandpass (typically the F814W bandpass).  Then, we masked regions
around detected objects and detected objects in the unmasked regions of images
of a longer-wavelength bandpass (typically the F160W or $H$ bandpass).
Finally, we repeated these steps for remaining images of longer-wavelength
bandpasses (typically the F222M or $K$ bandpass).  In this way, the
shorter-wavelength bandpasses (which are generally of higher sensitivity and
resolution but which are unsuitable for detecting high-redshift galaxies) were
used to detect low- and moderate-redshift galaxies and the longer-wavelength
bandpasses (which are generally of lower sensitivity and resolution but which
are suitable for detecting high-redshift galaxies) were used to detect
high-redshift galaxies.  Details of the object detection are summarized in
Table 2.

  We performed photometry using our quasi-optimal photometry technique, which
fits model spatial profiles of the detected objects to the space- and
ground-based images (Yahata et al.\ 2000).  We determined model spatial
profiles by using non-negative least squares image reconstruction (Puetter \&
Yahil 1999) of one or more of the space-based images.  The quasi-optimal
photometry technique offers three important advantages in comparison with
conventional methods:  (1) it provides higher signal-to-noise ratio
measurements, (2) it accounts for point-spread function variations between the
images, and (3) it accounts for uncertainty correlations between nearby,
overlapping neighbors.

\subsection{Photometric Redshift Measurements}
 
  Details of our photometric redshift measurements of faint galaxies in the HDF
and HDF--S WFPC2 and NICMOS fields have been described previously by Lanzetta
et al.\ (1996), Fern\'andez-Soto et al.\ (1999), and Yahata et al.\ (2000).  In
this section, we summarize important aspects of our photometric redshift
measurement techniques.

  We measured photometric redshifts (and spectral types) of galaxies using our
redshift likelihood technique (Lanzetta et al.\ 1996; Fern\'andez-Soto et al.\
1999) with a sequence of six spectrophotometric templates, including templates
of E/S0, Sbc, Scd, and Irr galaxies and low- and high-extinction starburst
galaxies (which we designate as SB1 and SB2), and incorporating the effects of
intrinsic and intervening absorption by neutral hydrogen.

  We assessed the accuracy and reliability of the photometric redshift
measurements by comparing photometric redshift measurements with spectroscopic
redshift measurements of galaxies identified in the HDF and HDF--S.  Results
indicate that at all redshifts $z < 6$ that have yet been examined, the
photometric redshift measurements are characterized by an RMS relative
dispersion with respect to the spectroscopic redshift measurements of $\Delta
z/(1 + z) \apl 0.065$ and that there are no known examples of photometric
redshift measurements that differ from spectroscopic redshift measurements by
more than a few times the RMS relative dispersion (Yahata et al.\ 2000;
Fern\'andez-Soto et al.\ 2001).  These results set an upper limit to the {\em
systematic} uncertainty of the photometric redshift technique, to which must be
added, of course, the uncertainty due to effects of photometric error to
determine the total uncertainty of a particular measurement.

\section{ANGULAR AREA VERSUS DEPTH RELATION}

  The angular area versus depth relations set the angular area accessible to
the observations as a function of ``depth,'' where ``depth'' might in general
be any one of a number of observed or intrinsic properties of galaxies (e.g.,
energy flux in a given observed-frame bandpass, luminosity at a given
rest-frame wavelength, or surface brightness at a given rest-frame wavelength).
For the purpose of the present analysis, we take ``depth'' to be unobscured
star formation rate intensity $x$.  The angular area versus depth relation must
account for several important effects:  First, the sensitivities of the
individual images (especially the NICMOS images) vary with position.  Second,
objects are detected in one or more observed-frame bandpasses, whereas results
are determined with respect to a single observed-frame bandpass or rest-frame
wavelength.  Finally, object detection is performed at multiple bandpasses,
which introduces complicated conditional selection criteria into the analysis.

  We determined the angular area versus depth relation as a function of
redshift $z$ and unobscured star formation rate intensity $x$ using surface
brightness sensitivity maps of the individual HDF and HDF--S WFPC2 and NICMOS
images.  First, we determined the surface brightness sensitivity maps by
scaling the sky variance maps in such a way as to match the faint envelopes of
surface brightnesses actually included into the segmentation maps of the
galaxies.  (This assumes that the sky variance maps trace the shape but not the
normalization of the surface brightness sensitivity maps.)  Then, we calculated
the surface brightness expected for given values of $z$ and $x$, using the $K$
correction of an Irr galaxy.  (No cosmological model is required, because the
dependence of surface brightness on redshift is independent of cosmological
model.)  Then, we summed the angular areas of the pixels that are sensitive
enough to detect the surface brightness expected for the given values of $z$
and $x$, which we identified by examining the surface brightness sensitivity
maps of the images used for object detection.  Finally, we repeated these steps
for a range of values of $z$ and $x$.

  The angular area versus depth relation $\Omega(z,x)$ as a function of
redshift $z$ and star formation rate intensity $x$ is shown in Figure 1.  In
Figure 1, white regions represent $\Omega(z,x) = 0$, black regions represent
$\Omega(z,x) = 11.7$ arcmin$^2$ (i.e.\ the total angular area spanned by the
observations), and grey regions represent intermediate values of $\Omega(z,x)$.

\section{STONY BROOK FAINT GALAXY REDSHIFT SURVEY}

  Our catalogs of photometry and photometric redshift measurements include
nearly 5000 faint galaxies, of which nearly 1000 yield redshift measurements $z
> 2$ and more than 50 yield redshift measurements $z > 5$ (ranging up to and
beyond $z = 10$).  The catalogs of photometry and photometric redshift
measurements and the angular area versus depth relations together constitute a
galaxy redshift survey to the faintest limits and highest redshifts yet
accessible, which we designate as the Stony Brook faint galaxy redshift survey.

  Properties of the survey are summarized in Figure 2, which shows the redshift
measurement distributions of all galaxies identified in the HDF and HDF--S
WFPC2 and NICMOS fields.  The redshift measurement distributions of galaxies
identified in the HDF and HDF--S WFPC2 field are characterized by broad peaks
at redshift $z \approx 1$ and tails extending to redshifts $z > 5$.  (The
distributions are statistically different from one another, with the HDF--S
WFPC2 field exhibiting a statistically significant excess of galaxies of
redshift $z > 2$ compared with the HDF, and both distributions exhibit
statistically significant large-scale fluctuations.  Because the HDF--S fields
were chosen due to their proximity to a known QSO of redshift $z = 2.2$, the
HDF--S field may be biased toward galaxies of redshift $z \approx 2$.)  The
redshift measurement distribution of galaxies identified in the HDF--S NICMOS
field is characterized by a broad peak at redshift $z \approx 1$ and a tail
extending to redshifts $z > 10$.

  We note several points about the nature of the galaxies of redshift
measurement $z > 6$ identified in our survey.  First, our photometric redshift
measurements have been directly compared with spectroscopic redshift
measurements at redshifts $z \apl 6$ but have not been directly compared with
spectroscopic redshift measurements at redshifts $z \apg 6$.  (See
Fern\'andez-Soto et al.\ 2001 for the latest description of this comparison and
for an assessment of the accuracy and reliability of the photometric and
spectroscopic redshift measurement techniques.)  Second, the redshift
likelihood functions of most (but not all) of the galaxies of redshift
measurement $z > 6$ are sufficiently broad or multiply-peaked that
lower-redshift solutions cannot be excluded.  Hence a ``best-fit'' redshift
measurement $z > 6$ does not by itself indicate a redshift $z > 6$---in every
case the redshift likelihood function must be consulted to identify the range
of allowed redshifts.  And third, our bootstrap error analysis (described
below) explicity accounts for cases of broad or multiply-peaked redshift
likelihood functions.  The emphasis of the present analysis is on redshifts
$z < 6$, and the galaxies of redshift measurement $z > 6$ do not bear
significantly on our primary conclusions.

\section{STAR FORMATION RATE INTENSITY DISTRIBUTION FUNCTION MEASUREMENTS}

  The objective of the analysis is to determine the star formation rate
intensity distribution function by considering all pixels in all galaxies in
all fields on an individual pixel-by-pixel basis.  In this section, we describe
our measurements of the star formation rate intensity distribution function.

  First, we determined the rest-frame 1500 \AA\ luminosity of each pixel, using
the measured energy flux of the pixel, the redshift of the pixel (which is set
by the redshift of the host galaxy), the empirical $K$ correction of the pixel
(which is set by the spectrophotometric template of the host galaxy), and the
cosmological model.  Then, we determined the star formation rate of each pixel,
using a Salpeter stellar initial mass function to scale rest-frame 1500 \AA\
luminosity to star formation rate.  Then, we determined the proper area of each
pixel, using the redshift of the pixel, the angular plate scale of the image,
and the cosmological model.  Then, we divided the star formation rate by the
proper area of each pixel to determine the star formation rate intensity $x$ of
the pixel.  The star formation rate intensity $x$ is measured, say, in units
of $M_\odot$ yr$^{-1}$ kpc$^{-2}$.

  Next, we summed the proper areas of all pixels within given star formation
rate intensity and redshift intervals.  Then, we determined the comoving
volumes that correspond to the redshift intervals, using the angular area
versus depth relation and the cosmological model.  Then, we divided the summed
proper areas by the appropriate star formation rate intensity intervals and
comoving volumes to determine the star formation rate intensity distribution
function $h(x)$ at values of the star formation rate intensity $x$ down to the
noise thresholds of the images.  Finally, we {\em excluded} measurements of
$h(x)$ at values of $x$ less than the values that could have been detected at
the $5 \sigma$ level over 95\% of the angular area of the survey.  The
segmentation maps that define the isophotal apertures of the galaxies contain,
of course, pixels of insignificant or even negative measured energy flux
(because the segmentation maps are defined with respect to smoothed versions
of the images.)  Thus low-intensity pixels near the edges of the segmentation
maps are unsuitable for incorporation into the measurement of $h(x)$, since
roughly half of these pixels have negative measured energy flux and so cannot
bear on the analysis.  Considering only high-intensity values of $x$
eliminates the possibility that low-intensity pixels near the edges of the
segmentation maps bias the measurement of $h(x)$.  We verified that the
measurement of $h(x)$ at our adopted star formation rate intensity thresholds
is not affected by pixel-to-pixel noise by redetermining $h(x)$ from within
the segmentation maps but with the actual images replaced by Gaussian noise
(according to the actual noise properties of the images).  The star formation
rate intensity distribution function $h(x)$ is measured, say, in units of
kpc$^2$ ($M_\odot$ yr$^{-1}$ kpc$^{-2}$)$^{-1}$ Mpc$^{-3}$.

  We determined measurement uncertainties of the star formation rate intensity
distribution function $h(x)$ by applying a variation of the ``bootstrap''
resampling technique described previously by Lanzetta, Fern\'andez-Soto, \&
Yahil (1998).  The bootstrap resampling technique explicitly accounts for
effects of sampling uncertainty, photometric uncertainty, and variance with
respect to the spectrophotometric templates.  First, we resampled objects from
the catalogs (allowing the possibility of duplication), added random noise to
the photometry (according to the actual noise properties of the images) and
redetermined the photometric redshift measurements, and added random noise to
the photometric redshift measurements (according to the actual noise properties
of the photometric redshift technique, as determined from the comparison of
photometric and spectroscopic redshift measurements).  Then, we determined the
star formation rate intensity distribution function $h(x)$ from the resampled,
perturbed catalogs of photometry and photometric redshifts exactly as from the
actual catalogs of photometry and photometric redshifts.  Finally, we repeated
these steps a large number of times in order to trace out the distributions of
measurement uncertainties.

\section{RESULTS}

  Results of our analysis are shown in Figure 3, which plots the star formation
rate intensity distribution function $h(x)$ versus the star formation rate
intensity $x$ as a function of redshift $z$.  Several points are evident on the
basis of Figure 3:

  First, the star formation rate intensity threshold of the observations is a
steep function of redshift, ranging from $x \approx 5 \times 10^{-4}$ $M_\odot$
yr$^{-1}$ kpc$^{-2}$ at $z = 0.5$, to $\approx 0.3$ $M_\odot$ yr$^{-1}$
kpc$^{-2}$ at $z = 5$, to $\approx 1$ $M_\odot$ yr$^{-1}$ kpc$^{-2}$ at $z =
10$.

  Second, at the redshifts $z \apl 1$ at which both low- and high-intensity
star formation rate intensities can be measured, $\log h(x)$ versus $\log x$ is
characterized by a relatively shallow dependence at $x \apl 10^{-2} \ M_\odot$
yr$^{-1}$ kpc$^{-2}$ and a relatively steep dependence at $x \apg 10^{-2} \
M_\odot$ yr$^{-1}$ kpc$^{-2}$.  The shallow dependence is sufficiently shallow
and the steep dependence is sufficiently steep that, from equation (1), the
bulk of the ultraviolet luminosity density is emitted where the two regions
join, i.e.\ at $x \approx 10^{-2} \ M_\odot$ yr$^{-1}$ kpc$^{-2}$.  This
implies that to measure the bulk of the ultraviolet luminosity density requires
sensitivity to star formation rate intensities {\em less than} the intensity $x
\approx 10^{-2} \ M_\odot$ yr$^{-1}$ kpc$^{-2}$ at which $h(x)$ steepens.  From
Figure 3, it is clear that the observations of the Hubble Deep Fields---which
are the most sensitive observations of faint galaxies ever obtained---can
detect the intensities at which $h(x)$ steepens only to redshift $z \approx
1.5$ or at most $z \approx 2$; at higher redshifts, the observations are not
sensitive to the intensities at which the bulk of the ultraviolet luminosity
density are emitted.

  Third, the high-intensity end of $h(x)$---which is directly measured at all
redshifts through $z \approx 10$---evolves more or less monotonically toward
higher values at higher redshifts.  For example, at $x \approx 3$ $M_\odot$
yr$^{-1}$ kpc$^{-2}$, $h(x)$ evolves from $h(x) \approx 10^{-5}$ $h_{100}$
kpc$^2$ ($M_\odot$ yr$^{-1}$ kpc$^{-2}$)$^{-1}$ Mpc$^{-3}$ at $z \approx 1.5$
through $\approx 10^{-4}$ $h_{100}$ kpc$^2$ ($M_\odot$ yr$^{-1}$
kpc$^{-2}$)$^{-1}$ Mpc$^{-3}$ at $z \approx 2.5$ through $\approx 10^{-3}$
$h_{100}$ kpc$^2$ ($M_\odot$ yr$^{-1}$ kpc$^{-2}$)$^{-1}$ Mpc$^{-3}$ at $z
\approx 4$.  There are no local (say $z < 1$) analogs of the very high
intensity (say $x > 1$ $M_\odot$ yr$^{-1}$ kpc$^{-2}$) regions that are seen at
higher redshifts, at least over the volumes probed by the Hubble Deep Fields.

  To quantitatively establish the effect of the star formation rate intensity
threshold of the observations on the determination of the star formation rate
density, we fitted $h(x)$ versus $x$ by a broken double power law model
\begin{equation}
h(x) = \left\{
\begin{array}{ll}
A \left( x/x_0 \right)^{-\alpha_1} & (x < x_0) \\
A \left( x/x_0 \right)^{-\alpha_2} & (x \ge x_0),
\end{array}
\right.
\end{equation}
which is described by slopes $\alpha_1$ and $\alpha_2$ at low and high
intensities, characteristic intensity $x_0$, and normalization $A$.  First, we
established fiducial parameters by fitting the model to the observations over
the redshift interval $z = 0.5$ to 1.  (These redshifts are low enough that the
observations are sensitive to low values of $x$ but not so low that redshift
measurement uncertainties can significantly affect measurements of luminosity
or star formation rate intensity.)  Results indicate a low-intensity slope
\begin{equation}
\alpha_1 = -1.44^{+0.22}_{-0.09},
\end{equation}
a high-intensity slope
\begin{equation}
\alpha_2 = -3.05^{+0.25}_{-0.33},
\end{equation}
and a characteristic intensity
\begin{equation}
\log x_0 =  -1.96 \pm 0.13 \ M_\odot \ {\rm yr} \ {\rm kpc}^2.
\end{equation}
The fiducial model is shown in Figure 3a.  Next, we fitted simple variants of
the fiducial model over other redshift intervals, fixing $\alpha_1$ and
$\alpha_2$ and either (1) scaling vertically (i.e.\ fixing $x_0$ and varying
$A$), (2) scaling horizontally (i.e.\ fixing $A$ and varying $x_0$), and (3)
scaling the break intensity [i.e.\ varying $x_0$ and $A$ in such a way that
$h(x)$ at low intensities does not change].  The variants of the fiducial
model are shown in Figure 3b.  Finally, we determined the star formation rate
density versus redshift by integrating over the models, to intensities as low
as the lowest star formation rate intensity $x = 6 \times 10^{-4}$ $M_\odot$
yr$^{-1}$ kpc$^{-2}$ observed in the local universe (Kennicutt 1998).

  Results are shown in Figure 4a, which plots the unobscured star formation
rate densities versus redshift obtained by integrating over the models.  For
comparison, Figure 4a also plots the unobscured star formation rate density
versus redshift obtained by Madau et al.\ (1996, 1998) by summing over the
observed galaxy energy fluxes.  From Figure 4a, it is evident that the
ultraviolet luminosity density obtained by integrating over the models
increases more or less monotonically with redshift, for each of the three
models.  Over the redshift interval $z = 0$ through 10, the increase ranges
from a factor 10 (for scaling the break intensity of the fiducial model)
through a factor 250 (for scaling the fiducial model vertically).  The
ultraviolet luminosity density obtained by Madau et al.\ (1996, 1998) by
summing over the observed galaxy energy fluxes significantly underestimates the
unobscured star formation rate density obtained by integrating over the models
at redshifts larger than $z \approx 2$.  The increase of the ultraviolet
luminosity density with redshift is a direct consequence of the monotonic
increase of the incidence of the highest intensity star forming regions with
redshift and must hold unless the actual functional form of $h(x)$ at redshifts
$z \apg 2$ differs dramatically from the functional forms of the three simple
variants of the fiducial model.  We conclude that the unobscured star formation
rate density---or equivalently the ultraviolet luminosity density---plausibly
increases monotonically with redshift through the highest redshifts observed.

  But it also appears from Figure 4a that our measurements and the measurements
of Madau et al.\ (1996, 1998) of the unobscured star formation rate density are
discordant at redshifts $z = 0$ to 2.  This is a consequence not of
cosmological surface brightness dimming effects but rather of a mismatch of the
rest-frame wavelengths from which the star formation rate densities were
measured.  We measured star formation rate densities by scaling from measured
luminosity densities at rest-frame wavelength 1500 \AA\ at all redshifts.  In
contrast, Madau et al.\ (1996, 1998) measured star formation rate densities by
scaling from measured luminosity densities at rest-frame wavelength 2800 \AA\
at redshifts $z < 2$ (based on results of Lilly et al.\ 1996 and Connolly et
al.\ 1997) and at rest-frame wavelength 1500 \AA\ at redshifts $z > 2$ (based
on results from the Hubble Deep Field), adopting a {\em flat} $f_\nu$ spectrum
to scale luminosity density to star formation rate density.  Thus our
measurements and the measurements of Madau et al.\ (1996, 1998) should be
consistent only if the rest-frame ultraviolet luminosity density of the
universe exhibits a flat $f_\nu$ spectrum at redshifts $z = 0$ to 2.

  To test this possibility, we measured the spectrum of the luminosity density
versus redshift, based on galaxies observed in the HDF and HDF--S using our
catalog of photometry and photometric redshifts.  Results are shown in Figure
5.  The ratio of the luminosity density at rest-frame wavelength 2800 \AA\ to
the luminosity density at rest-frame wavelength 1500 \AA\ is $\approx 2$ at
redshifts $z < 2$.  In other words, the flat $f_\nu$ spectrum adopted by Madau
et al.\ (1996, 1998) is {\em inconsistent} with the measured spectrum of the
luminosity density at redshifts $z < 2$.  Consequently, the measurements of
Madau et al.\ (1996, 1998) at redshifts $z < 2$ must be {\em reduced} by a
factor $\approx 2$ to bring them into consistency with measurements of the star
formation rate density based on rest-frame 1500 \AA\ luminosity density.  The
results corrected in this way are shown in Figure 4b.  It appears from Figure
4b that our measurements and the corrected measurements of Madau et al.\ (1996,
1998) of the unobscured star formation rate density are consistent at redshifts
$z < 2$.

  Adoption of the currently favored non-zero cosmological constant cosmological
model of vaccuum energy density $\Omega_\Lambda = 0.7$ and matter density
$\Omega_M = 0.3$ does not qualitatively change our conclusions.  Under this
cosmological model, the values of the unobscured star formation densities are
reduced, by amounts that range from $\approx 20\%$ for the lowest-redshift
point to $\approx 50\%$ for the highest-redshift point.

\section{CONNECTION WITH THE NEUTRAL HYDROGEN COLUMN DENSITY DISTRIBUTION
FUNCTION}

  The star formation rate intensity distribution function $h(x)$ is exactly
analogous to the column density distribution function $f(N)$ (as a function of
neutral hydrogen column density $N$) measured from damped \lya\ absorption
systems detetected toward high-redshift background QSOs in that $h(x) dx$ and
$f(N) dN$ both measure incidence per length of cosmologically distributed
material.  The connection between $h(x)$ and $f(N)$ can be made explict by
adopting a ``Kennicutt'' (1998) relation between star formation rate intensity
$x$ and neutral hydrogen column density $N$
\begin{equation}
x = 8.9 \times 10^{-5} \left( \frac{N}{1.2 \times 10^{20} \ \cmjj}
\right)^{1.4} \ M_\odot \ {\rm yr}^{-1} \ {\rm kpc}^{-2}.
\end{equation}
(Kennicutt 1998 multiplied observed energy fluxes by a factor 2.8 to correct
for effects of obscuration by intervening dust.  Here we divide the energy
fluxes of Kennicutt 1998 by this factor 2.8 to {\em uncorrect} for effects of
obscuration by intervening dust, i.e.\ to establish a relation between {\em
unobscured} star formation rate intensity $x$ and gas column density $N$.  We
consider neutral hydrogen column density to represent gas column density
because the molecular content of high column density QSO absorption systems is
measured to be low.) Then $h(x)$ and $f(N)$ are related as
\begin{equation}
h(x) \; dx \; dl = f(N) \; dN \; dX,
\end{equation}
where $dl$ is a comoving length element and $dX$ is an absorption distance
element.  We applied equations (6) and (7) to express measurements of $f(N)$ at
redshifts $z \approx 1.6$ through 5 (Lanzetta et al.\ 1991; Storrie-Lombardi,
Irwin, \& McMahon 1996) in terms of $h(x)$.

  Results are shown in Figure 3b.  The amplitude of the measurements of $f(N)$
at redshifts $z \approx 1.6$ through 5 expressed in terms of $h(x)$ match the
amplitude of the measurements of $h(x)$.  Further, application of equation (5)
to a power-law form of the column density distribution function $f(N) \propto
N^{-\beta}$ of slope $\beta = 1.48 \pm 0.30$ (Storrie-Lombardi, Irwin, \&
McMahon 1996) predicts a power-law form of the star formation rate intensity
distribution function $h(x) \propto x^{-\alpha_1}$ of slope $\alpha_1 = 1.34
\pm 0.22$, which is in excellent agreement with the slope $\alpha_1 =
1.44^{+0.22}_{-0.09}$ measured at redshifts $z = 0.5$ to $1$.  We take the
remarkable agreement between $f(N)$ expressed in terms of $h(x)$ and
$h(x)$---in terms of both amplitude and slope---to (1) suggest that the actual
functional form of $h(x)$ at redshifts $z \apg 2$ does not differ dramatically
from the function forms of the three simple variants of the fiducial model (and
that the third variant is slightly favored over the other two variants) and (2)
establish a quantitative connection between high-redshift galaxies and high
column density gas and suggest that high-redshift damped \lya\ absorption
systems trace lower star formation rate intensity regions of the same galaxies
detected in star light in the HDF and HDF--S.

  Of course, $f(N)$ and $h(x)$ are measured on very different physical scales.
Specifically, $f(N)$ is measured on a physical scale of $\approx 1$ pc (which
is set by the size of the continuum emitting regions of QSOs), whereas $h(x)$
is measured on a physical scale of $\approx 200$ pc (which is set by the
angular resolution of the HDF and HDF--S images).  The remarkable agreement
between $f(N)$ expressed in terms of $h(x)$ and $h(x)$ nevertheless establishes
that faint galaxies in the HDF and HDF--S obey a {\em statistical} Kennicutt
(1998) relation that is consistent with the Kennicutt (1998) relation of nearby
galaxies.  In other words, the relationship between column density measured on
a scale of $\approx 1$ pc and star formation rate intensity measured on a scale
of $\approx 200$ pc for faint galaxies in the HDF and HDF--S is {\em on
average} consistent with the relationship between column density and star
formation rate intensity measured on galactic scales for nearby galaxies.

\section{SUMMARY AND CONCLUSIONS} 

  We address the effects of cosmological surface brightness dimming on
observations of faint galaxies by examining the distribution of ``unobscured''
star formation rate intensities versus redshift, which we use to assess the
ultraviolet luminosity density versus redshift, using our photometry and
photometric redshift measurements of faint galaxies in the HDF and the HDF--S
WFPC2 and NICMOS fields.  We find the following results:
\begin{enumerate}

  \item Previous measurements have missed a dominant fraction of the ultraviolet
luminosity density of the universe at high redshifts by neglecting cosmological
surface brightness dimming effects, which are important at redshifts larger
than $z \approx 2$.

  \item The incidence of the highest intensity star forming regions increases
monotonically with redshift.

  \item The ultraviolet luminosity density plausibly increases monotonically
with redshift through the highest redshifts observed.

  \item Previous measurements of the ultraviolet luminosity density at redshifts
$z < 2$ must be reduced by a factor $\approx 2$ to allow for the spectrum of 
the luminosity density between rest-frame wavelengths 1500 and 2800 \AA.

  \item The distribution of star formation rate intensities matches the
distribution of neutral hydrogen column densities at redshifts $z \approx 2$
through 5, which establishes a quantitative connection between high-redshift
galaxies and high column density gas and suggests that high-redshift damped
\lya\ absorption systems trace lower star formation rate intensity regions of
the same galaxies detected in star light in the HDF and HDF--S.

\end{enumerate}
Because our measurements neglect the effects of obscuration by dust, they
represent {\em lower limits} to the total star formation rate density.  Our
analysis suggest that star formation in the very early universe may have
occurred at a much higher rate than is generally believed and that cosmological
surface brightness dimming effects cannot be ignored when interpreting
statistical properties of the high-redshift galaxy population.

\acknowledgements

This research was supported by NASA grant NAGW--4422 and NSF grant
AST--9624216.  AFS was supported by a European Union Marie Curie Fellowship.
The authors acknowledge helpful comments from an anonymous referee.

\newpage

\begin{center}
\begin{tabular}{p{1.75in}cc}
\multicolumn{3}{c}{TABLE 1} \\
\multicolumn{3}{c}{DETAILS OF THE OBSERVATIONS} \\
\hline
\hline
\multicolumn{1}{c}{Field} & \multicolumn{1}{c}{Bandpass} &
\multicolumn{1}{c}{Reference} \\
\hline
HDF \dotfill           & F300W & 1 \\
                       & F450W & 1 \\
                       & F606W & 1 \\
                       & F814W & 1,2 \\
                       & F110W & 3,4 \\
                       & F160W & 3,4 \\
                       & $J$   & 5 \\
                       & $H$   & 5 \\
                       & $K$   & 5 \\
HDF--S WFPC2 \dotfill  & F300W & 6 \\                                                                  & F450W & 6 \\
                       & F606W & 6 \\
                       & F814W & 6 \\
                       & $U$   & 7 \\
                       & $B$   & 7 \\
                       & $V$   & 7 \\
                       & $R$   & 7 \\
                       & $I$   & 7 \\
                       & $J$   & 7 \\
                       & $H$   & 7 \\
                       & $K$   & 7 \\
HDF--S NICMOS \dotfill & F110W & 8 \\
                       & F160W & 8 \\
                       & F222M & 8 \\
                       & $U$   & 9 \\
                       & $B$   & 9 \\
                       & $V$   & 9 \\
                       & $R$   & 9 \\
                       & $I$   & 9 \\
\hline
\end{tabular}
\parbox{3.25 in}{REFERENCES---(1) Williams et al.\ (1996); (2) Gilliland,
Nugent, \& Phillips (1999); (3) Thompson et al.\ (1999); (4) Dickinson et al.\
(2000); (5) Dickinson et al.\ (2001), in preparation; (6) Casertano et al.\
(2000); (7) da Costa et al.\ (2001); (8) Williams et al.\ (2000); (9) ESO
VLT--UT1 Science Verification (1998).}
\end{center}
 
\newpage
 
\begin{center}
\begin{tabular}{p{1.75in}l}
\multicolumn{2}{c}{TABLE 2} \\
\multicolumn{2}{c}{DETAILS OF THE OBJECT DETECTION} \\
\hline
\hline
\multicolumn{1}{c}{Field} & \multicolumn{1}{c}{Object Detection Bandpasses} \\
\hline
HDF \dotfill           & F814W, F160W, $K$ \\
HDF--S WFPC2 \dotfill  & F814W, $H$, $K$ \\
HDF--S NICMOS \dotfill & F160W, F222M \\
\hline
\end{tabular}
\end{center}

\newpage

\figcaption{Angular area $\Omega$ accessible to the observations as a function
of logarithm of star formation rate intensity $x$ and redshift $z$.  Star
formation rate intensity $x$ is measured in units of $M_\odot$ yr${-1}$
kpc${^-2}$.  White represents $\Omega = 0$, black represents $\Omega = 11.7$
arcmin$^2$, and grey represents intermediate values of $\Omega$.}

\figcaption{Redshift measurement distributions of all galaxies identified in
the (a) HDF and HDF--S (b) WFPC2 and (c) NICMOS fields.}

\figcaption{Star formation rate intensity distribution function $h(x)$ versus
star formation rate intensity $x$ at redshifts ranging from $z = 0$ to 0.5
through $z = 6$ to 10.  Vertical error bars show $1 \sigma$ uncertainties
(determined by a bootstrap resampling technique), and horizontal error bars
show bin sizes.  (For many points, the vertical error bars are smaller than the
points.)  (a) Black circles show $h(x)$ determined from observations of faint
galaxies.  Green line segments show the best-fit broken double power law model
fitted to the observations at redshifts $z = 0.5$ to 1.  (b) Black circles show
$h(x)$ determined from observations of faint galaxies.  Magenta open circles
show $h(x)$ determined by expressing measurements of $f(N)$ in terms of $h(x)$.
Line segments show variants of the best-fit broken double power law model
obtained by scaling vertically (red), scaling horizontally (blue), and scaling
the break intensity (green).}

\figcaption{(a) Unobscured star formation rate densities $\dot{\rho}_S$ versus
redshift $z$.  Points plot unobscured star formation rate densities versus
redshift obtained by integrating over the models, fixing $\alpha_1$ and
$\alpha_2$ and either (1) scaling vertically (red squares), (2) scaling
horizontally (blue triangles), and (3) scaling the break intensity (green
circles) and the unobscured star formation rate density versus redshift
obtained by Madau et al.\ (1996, 1998) by summing over the observed galaxy energy fluxes (black open circles).  Vertical error bars show $1 \sigma$
uncertainties, which for the red squares, blue triangles, and green circles are
determined by a bootstrap resampling technique.  (b) Same as (a), except that
the measurements of Madau et al.\ (1996, 1998) at redshifts $z < 2$ are reduced
by a factor 2 to bring them into consistency with measurements of the star
formation rate density based on rest-frame 1500 \AA\ luminosity density.}

\figcaption{Ratio of the luminosity density at rest-frame wavelength 2800 \AA\
to the luminosity density at rest-frame wavelength 1500 \AA\ at redshifts $z <
10$, based on galaxies observed in the HDF and HDF--S using our catalog of
photometry and photometric redshifts.}

\newpage

\begin{figure}
\plotone{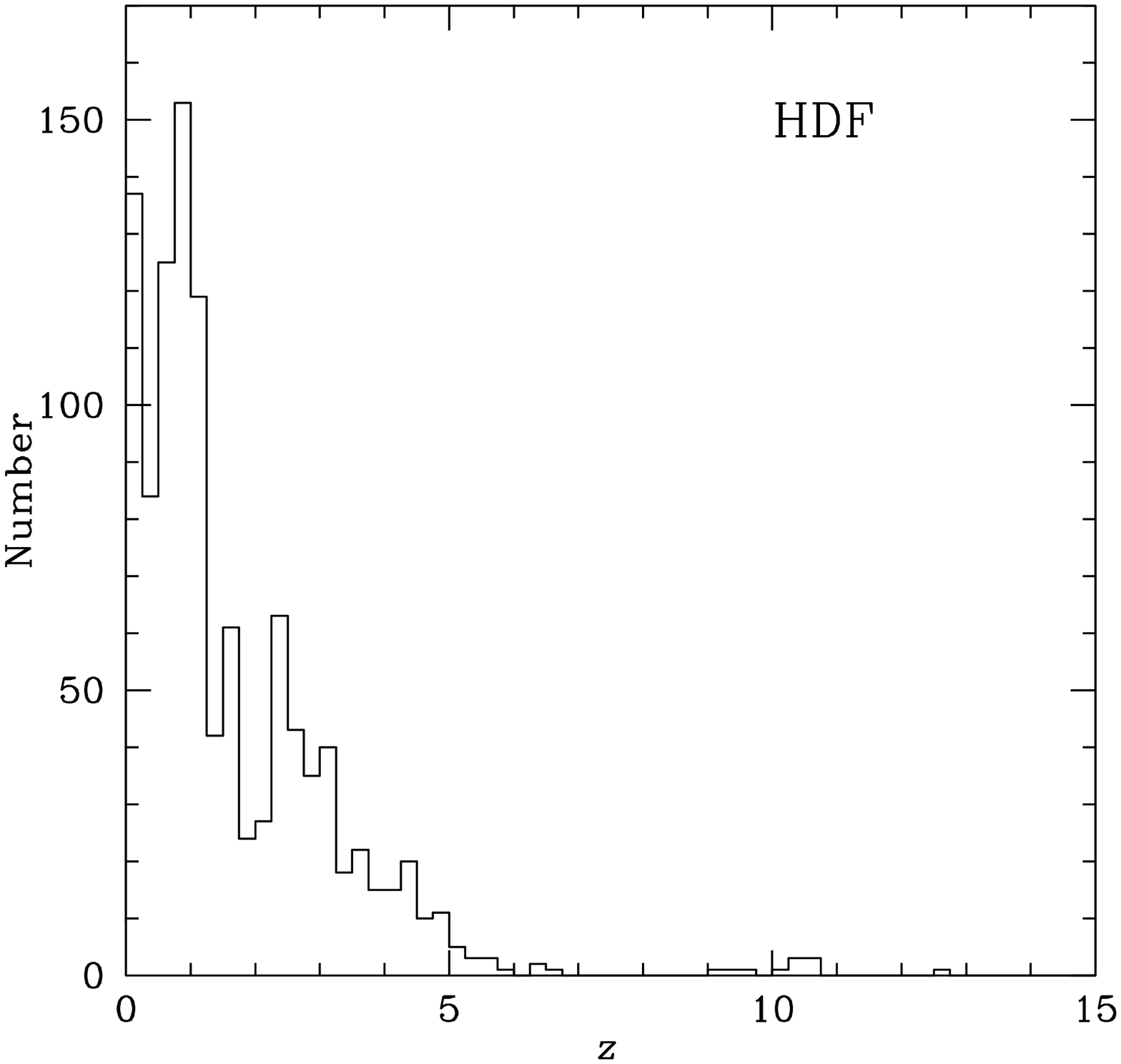}
\end{figure}

\begin{figure}
\plotone{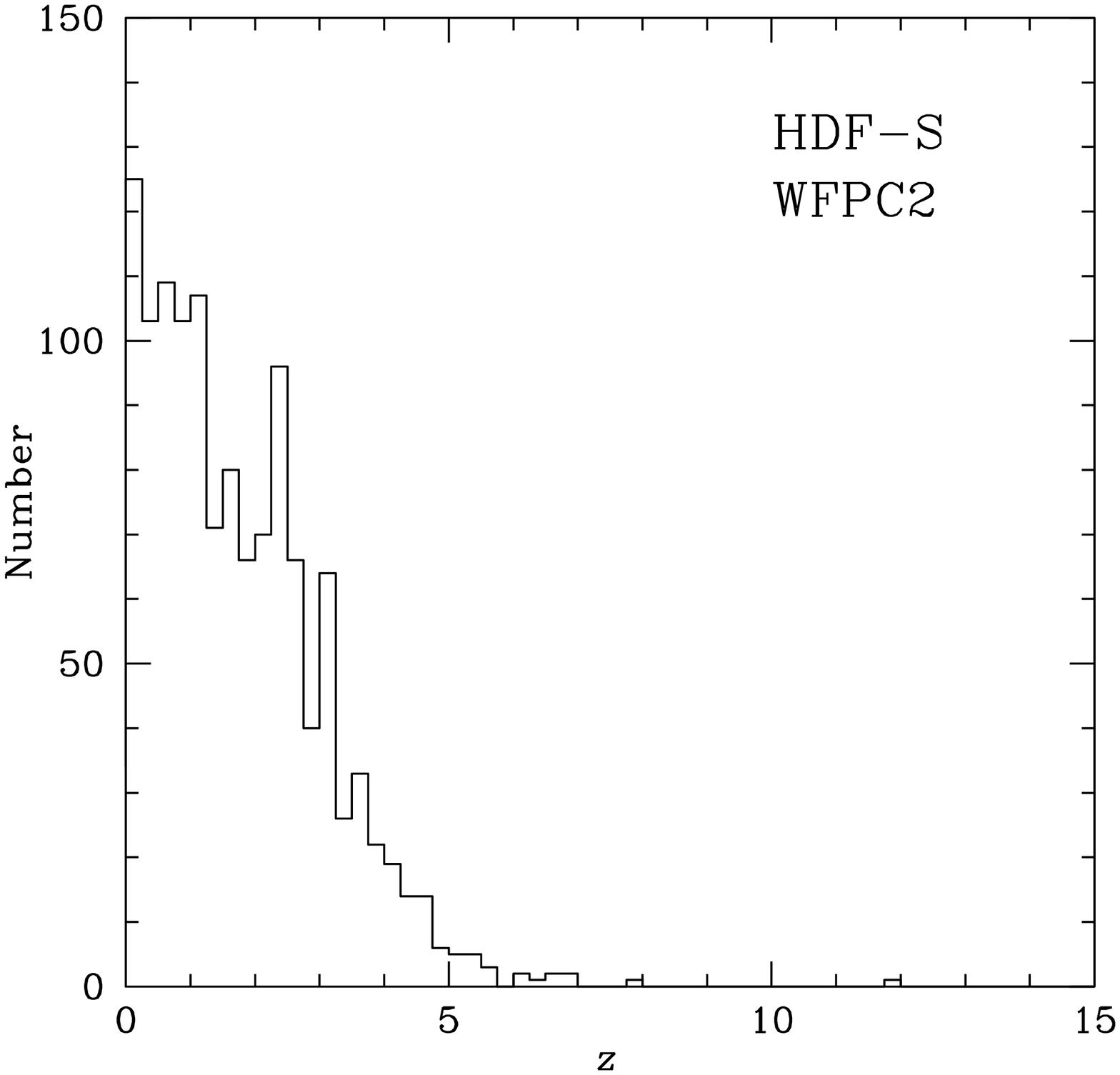}
\end{figure}

\begin{figure}
\plotone{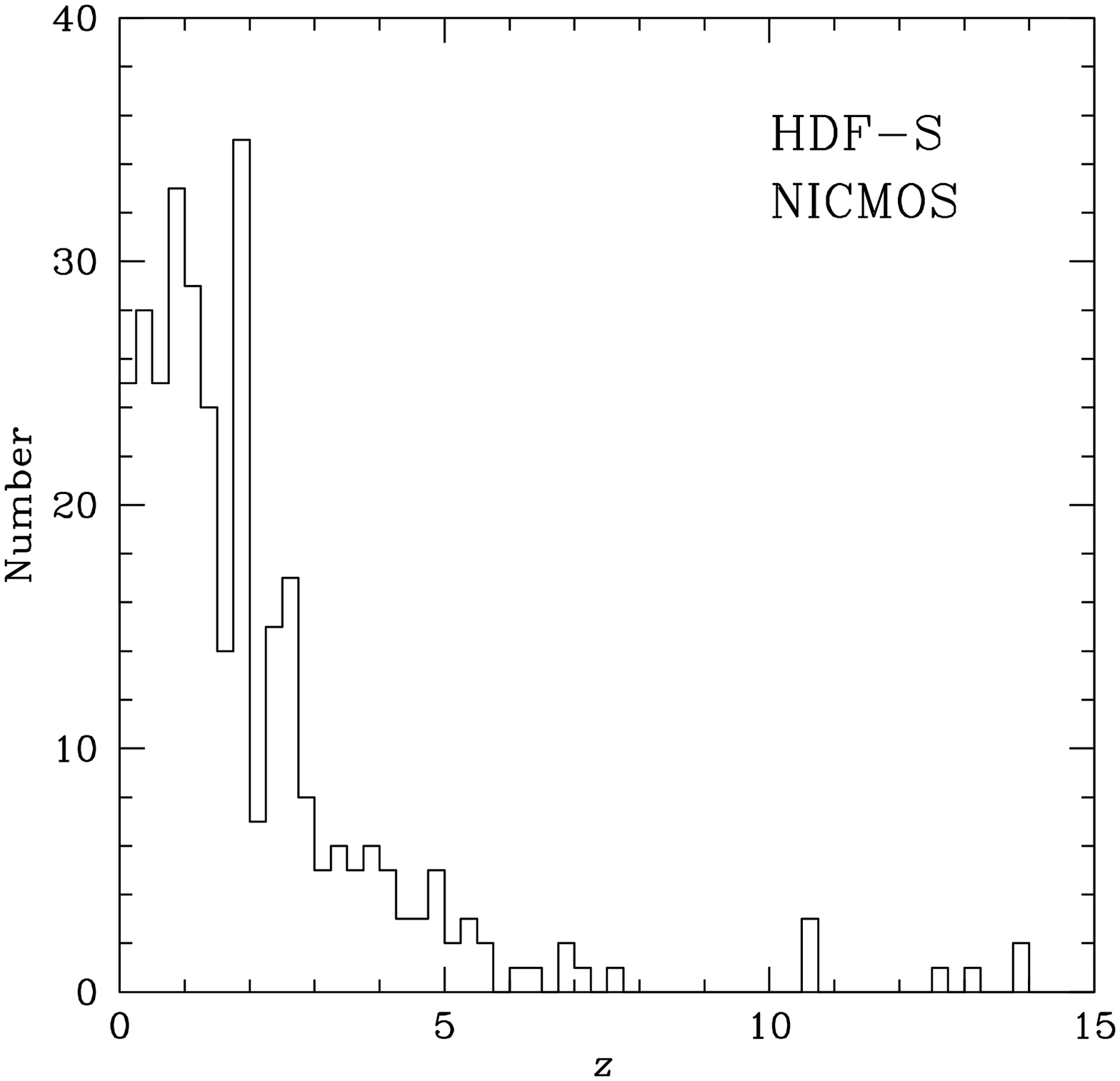}
\end{figure}

\begin{figure}
\plotone{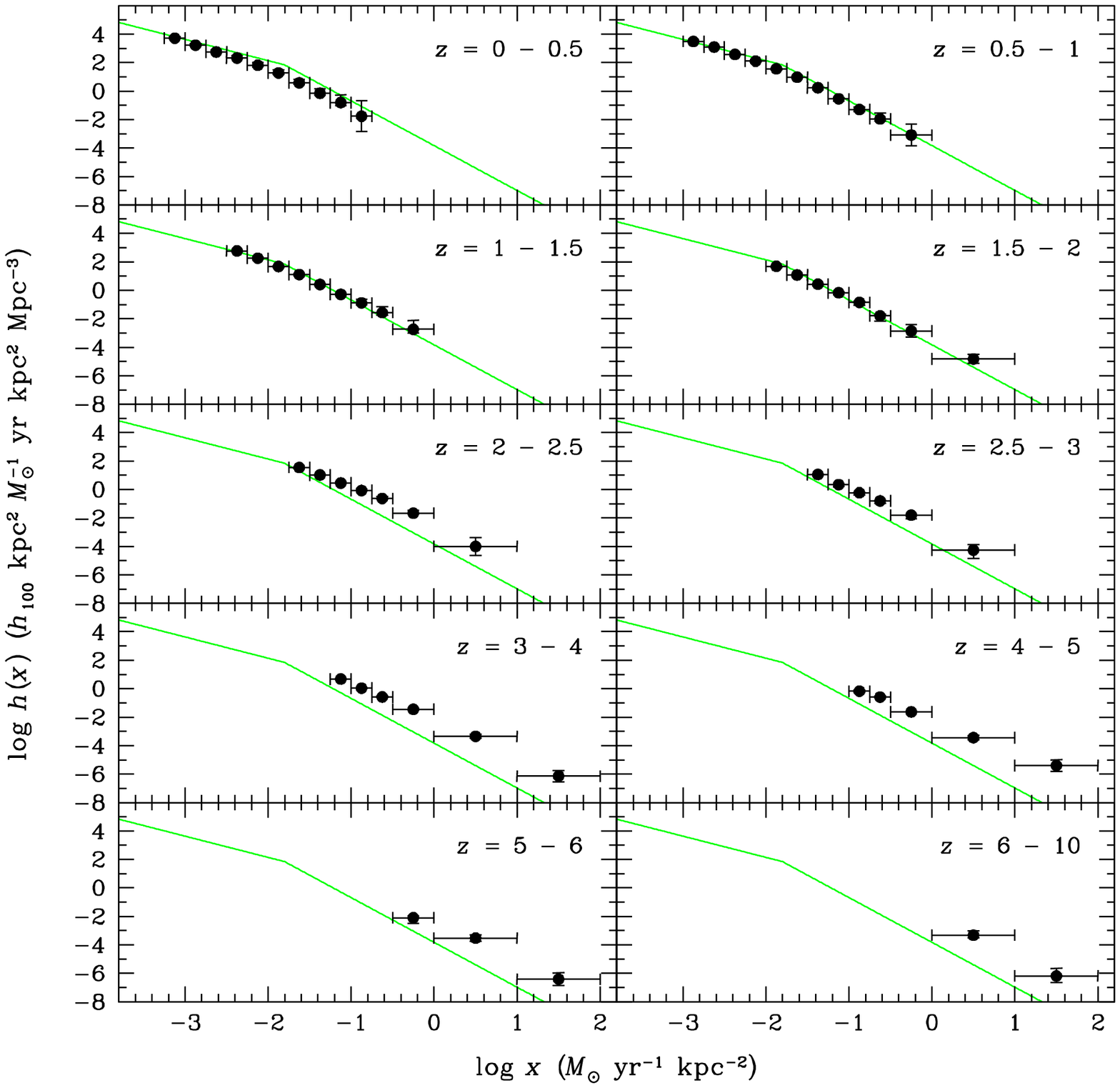}
\end{figure}

\begin{figure}
\plotone{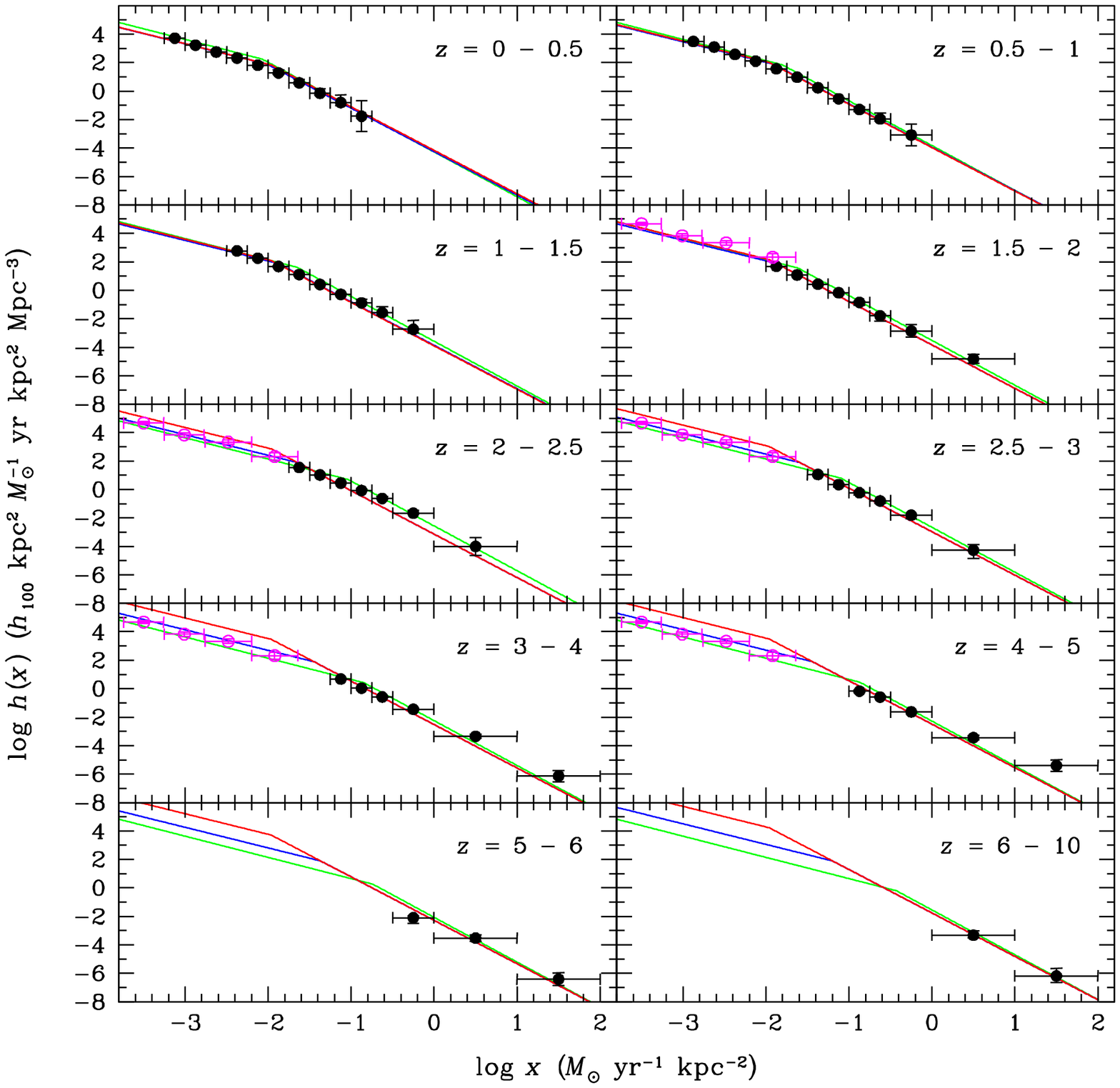}
\end{figure}

\begin{figure}
\plotone{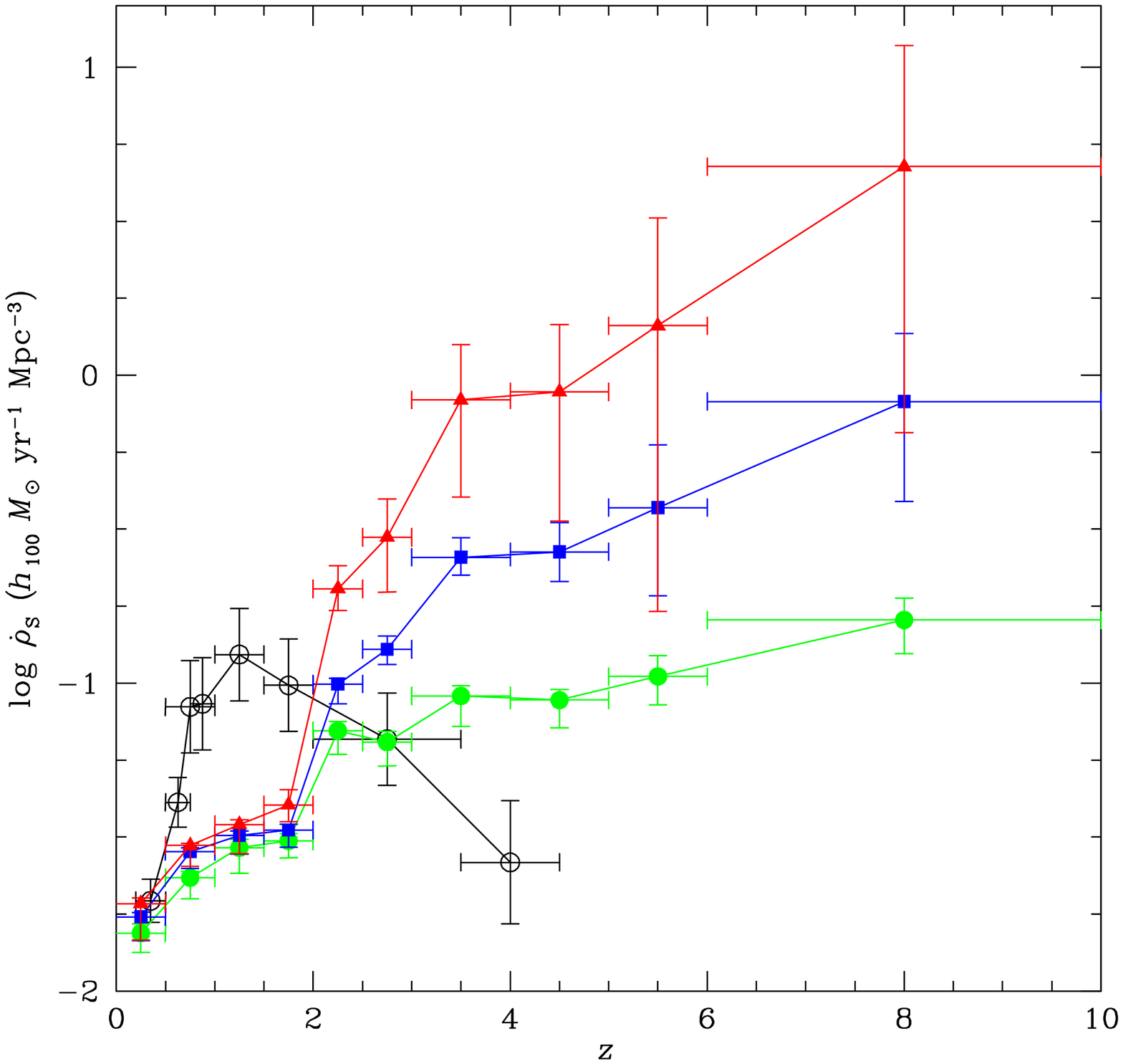}
\end{figure}

\begin{figure}
\plotone{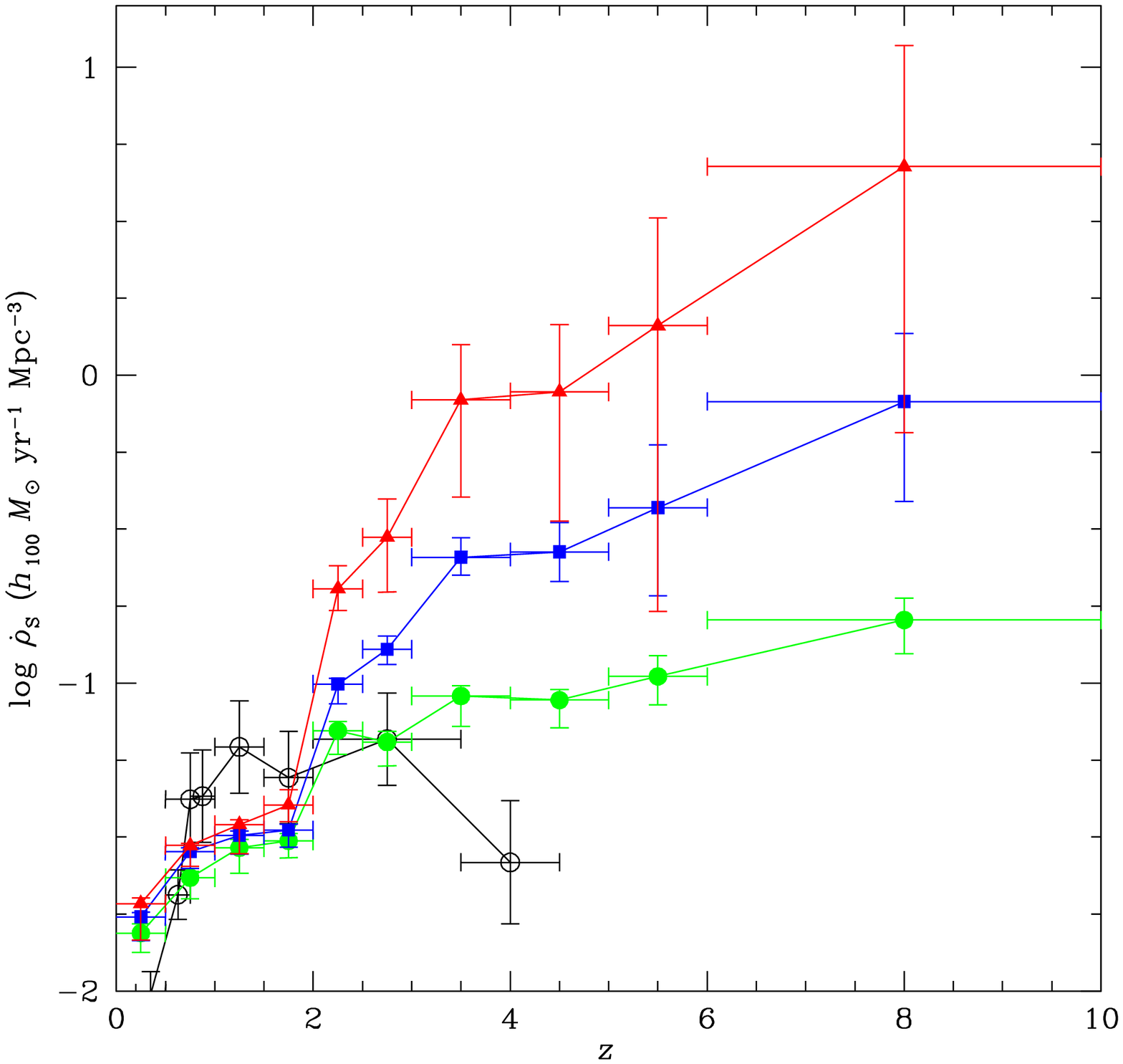}
\end{figure}

\begin{figure}
\plotone{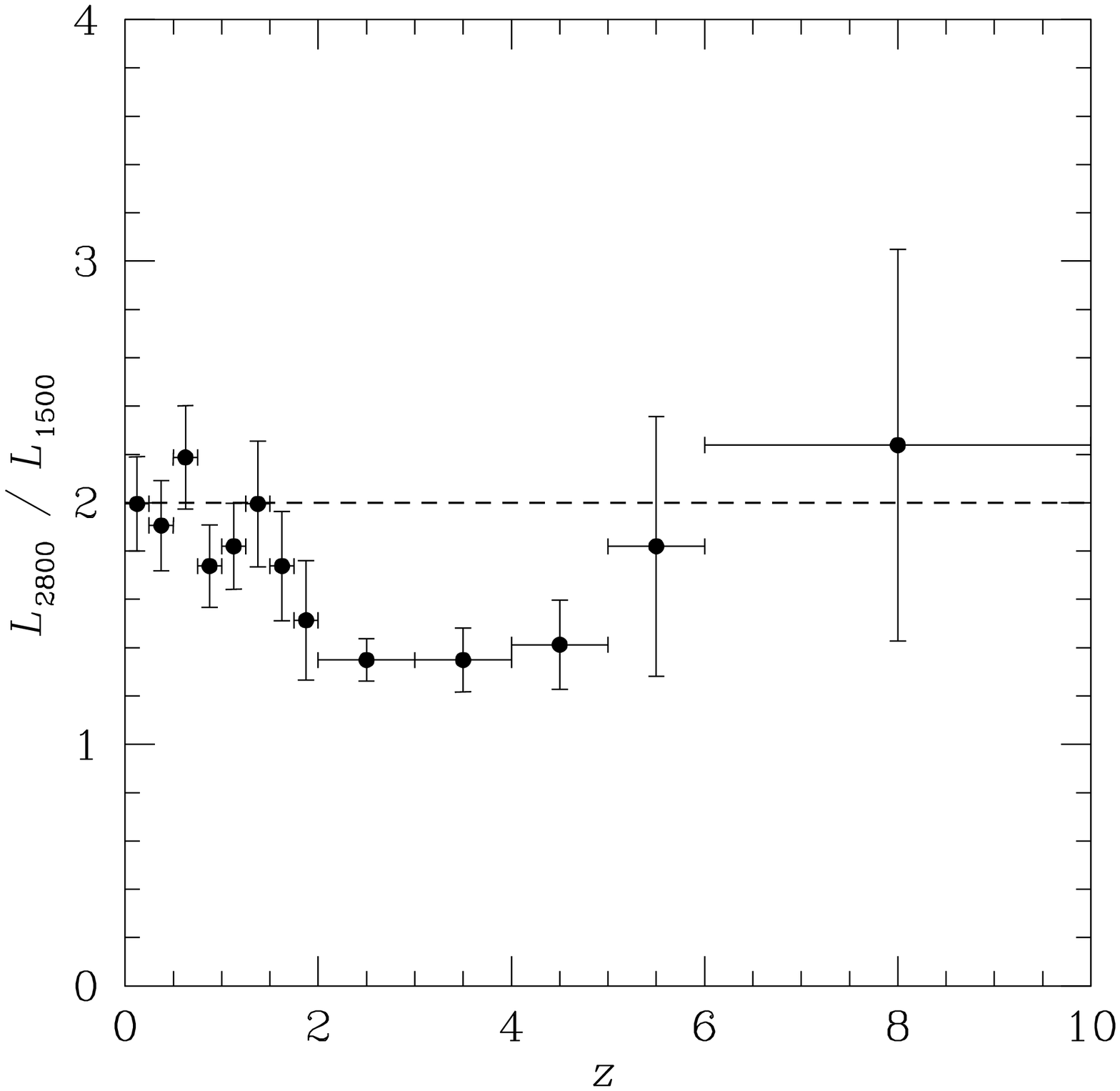}
\end{figure}

\end{document}